\documentclass{article}
\usepackage{graphicx}

\usepackage{amssymb,amsthm,amsmath}

\usepackage{xcolor,paralist,hyperref,titlesec,fancyhdr,etoolbox}
\usepackage[style=ieee, backend=biber]{biblatex}
\usepackage{enumitem}
\usepackage{authblk} 

\titleformat{\section}[block]{\normalfont\Large\bfseries}{\thesection}{1em}{}
\titlespacing*{\section}{0pt}{3.5ex plus 1ex minus .2ex}{2.3ex plus .2ex}

\hypersetup{ colorlinks=true, linkcolor=black, filecolor=black, urlcolor=black }

\usepackage{lipsum}

\title{On a Liquid Krypton TPC for double positron decay searches}

\author[1]{S. Torelli}
\author[2]{A. Simón Estévez}
\author[1,3]{F. Monrabal Capilla}

\affil[1]{\footnotesize Donostia International Physics Center (DIPC), Paseo Manuel de Lardizabal 4, E-20018 Donostia-San Sebastián, Spain}
\affil[2]{\footnotesize Instituto de Física Corpuscular (IFIC), CSIC-Universitat de València, C/ Catedrático José Beltrán 2, E-46980 Paterna, Valencia, Spain}
\affil[3]{\footnotesize Ikerbasque, Basque Foundation for Science, E-48009 Bilbao, Spain}

\date{}

\begin{document}
\maketitle

\begin{abstract}
Double positron decay features the emission of four 511 keV gamma rays from positron alongside the well-known 'double blob' signature which can be tracked with a TPC. Successfully detecting these gammas creates a unique signature that cannot be replicated by any background. This enables a virtually background-free search if the four gammas are efficiently tagged. In this work, we propose the concept of a liquid Krypton Time Projection Chamber designed to search for double positron decay using this four-gamma tagging concept, and we'll show how already a small version of this prototype could set the world best limit on the search of this decay towards a sensitivity of $10^{29}-10^{30}$ y with a ton scale detector.
\end{abstract} 
\section{Introduction}
The observation of double beta decay would reveal the fundamental nature of the neutrino, proving whether it is a Majorana or a Dirac particle. To search for this decay, various experiments utilize different technological approaches. These range from High-Purity Semiconductor Detectors \cite{Agostini_2020,Saleh:2026mon} and Cryogenic Bolometers \cite{Guardincerri:2011zz} to gaseous and liquid Xe (LXe) Time Projection Chambers (TPC) \cite{2026,EXO-200:2025obe}, with current sensitivities reaching the order of $10^{27}$ years. In this context High Purity Germanium (HPGe) detector offers unmatched energy resolution to cleanly isolate the signal peak, but scaling up its mass is slow and expensive due to the manufacturing process. Liquid Xenon TPCs provide massive exposure and excellent self-shielding due to high density, but suffer from poorer resolution, no topological discrimination and it's affected by the very high Xenon cost at present. Gaseous Xenon (GXe) TPCs trade density for tracking topology, allowing them to visually image the two-electron signal, despite the limited efficiency, and reject single electron backgrounds. Finally, Loaded Liquid Scintillators scale the fastest and cheapest in terms of mass, but their poor energy resolution prevents event-by-event discrimination, forcing them to rely heavily on statistical modeling against a large background.

Double positron decay, through its 0$\nu$ decay mode, may offer a unique opportunity to investigate whether neutrinos are Majorana particles. In this decay, in addition to the emission of two positrons, there is an exceptionally clear signature. This signature consists of four 511 keV gammas emitted in collinear pairs alongside the main double positron track with a total kinetic energy shared among the two positrons of $E_{kin} = Q_{val} -4\cdot \mathrm{m_e}$ \cite{10.3389/fphy.2019.00012} where $Q_{val}$ is the energy difference between the two nuclei and $\mathrm{m_e}$ is the mass of the electron. Detecting these 511 keV gammas together with a third track in the center would be the smoking gun for the observation this decay.  This detection requires a detector featuring a target with a density high enough to allow the gammas to separate from the main track yet sufficient to contain them, combined with strong 3D tracking capabilities.

This paper introduces the concept of a Liquid Krypton Time Projection Chamber (LKr TPC) as a detector for double positron decay searches and evaluates why Krypton can be an optimal target material for this application. The first section describes the detector concept, including the optimal choice of target material and suitable readout technologies. Furthermore, we evaluate the detector's sensitivity for the $0\nu\beta^+\beta^+$ and $2\nu\beta^+\beta^+$ decay processes, with particular emphasis on the expected background and demonstrating the exceptional rejection power provided by the highly distinct topological signatures.
\section{The detector concept}  
To search for double positron decay by exploiting its highly distinct topological signature, thus a Time Projection Chamber was selected for this study. A TPC, being an intrinsic 3D detector, provides three-dimensional event reconstruction within the active target volume, making it an excellent tool for identifying the $\beta^+\beta^+$ signature. Among the candidate isotopes for double positron decay evaluated in \cite{Barea_2013}, only $^{124}\text{Xe}$ and $^{78}\text{Kr}$ are noble gases suitable for operating a TPC. While both isotopes share a comparable $Q$-value (2.865 MeV and 2.846 MeV, respectively), Krypton offers distinct advantages. Its natural isotopic abundance is nearly four times higher than that of Xenon (0.355\% versus 0.0952\%). Furthermore, high commercial and scientific demand makes Xenon more than an order of magnitude more expensive than Krypton, reinforcing the selection of Krypton as the optimal target material.

A gaseous target was evaluated but soon ruled out because, even at a high pressure of $10\text{ bar}$, the interaction length for $511\text{ keV}$ photons is $1.21\text{ m}$ in Xenon and $2.28\text{ m}$ in Krypton \cite{matthew_newville_202X_xraydb}. Because the drift length and diameter of most TPCs for low-energy physics rarely exceed $\mathcal{O}(1\text{ m})$, employing a gas phase poses severe containment challenges and heavily limits detection efficiency. Consequently, a liquid target was selected to ensure proper event containment. In the liquid phase, the interaction lengths for $511\text{ keV}$ gamma rays are significantly reduced to $3.5\text{ cm}$ for Xenon and $5\text{ cm}$ for Krypton \cite{matthew_newville_202X_xraydb}, making a compact detector design feasible.
In both cases, containing the $511\text{ keV}$ photons within a large active detector volume is straightforward. However, the larger interaction length of liquid krypton offers a distinct advantage for cluster separation. Specifically, $67\%$ of the photons interact beyond a distance of $2\text{ cm}$ from their production vertex in $\text{LKr}$, compared to only $56\%$ in liquid xenon. This lower density of close interactions in $\text{LKr}$ prevents track overlapping and allows for higher track cluster separation, providing another point for its selection as the optimal target material.

The choice of the readout system is critical to the detector's performance, as it must provide both excellent spatial resolution and good energy resolution. To accurately resolve closely spaced tracks, the spatial resolution should match the 3 times the order of magnitude of the electron diffusion during drift. For a 50 cm drift length in liquid Xenon, the electron diffusion is estimated to be approximately 1.1 mm \cite{chen2011}. Because no experimental measurements of electron diffusion currently exist for liquid Krypton, the Xenon data from \cite{chen2011} serves as a reasonable proxy for LKr. Additionally, simulation-based diffusion predictions from \cite{Cao2026} indicate that the transverse electron diffusion in LKr is approximately a factor of two smaller than in LXe, while the longitudinal diffusion is even lower than the transverse component.
Based on these considerations, an excellent choice would be a single-phase TPC utilizing either a charge wire readout or a pixelated charge readout with a $\sim$3 mm pitch. Operating in a single phase eliminates the additional electron diffusion that occurs in a gaseous region, while avoiding the technical complexities associated with the maintaining of the double phase. Furthermore, a direct charge readout bypasses the complications of vacuum ultraviolet (VUV) light detection and prevents the spatial blurring inherent to light readouts, where isotropic ($4\pi$) photon emission induces signals across multiple neighboring sensors. 
Good examples of already developed and suitable readout technologies include GAMPix \cite{Shutt:2024che}, a low-noise pixelated charge readout that features a $500\ \mu\text{m}$ pixel pitch, and low noise levels that enable an energy resolution below $1\%$ at $511\text{ keV}$. Alternatively, the wire readout systems engineered for the DUNE experiment—which currently utilizes a $4.7\text{ mm}$ wire pitch—could provide a viable, high-density alternative. Another solution might be the LarPix developed for DUNE \cite{LarPix} with a $\sim$3 mm pitch and a low electronic noise. A photomultiplier system may be positioned behind the cathode to detect primary scintillation light, establishing the absolute $Z$-coordinate of the interaction vertex. Detecting primary scintillation for events above $300\text{ keV}$ is highly feasible; as demonstrated in \cite{5d1ffb2b2999521e412dda95}, a light collection efficiency of just $0.8\%$ is sufficient to reliably detect the primary scintillation light from a much lower energy ($41.5\text{ keV}$) event in gaseous Xenon. Current readout technologies already developed for liquid TPCs for DM and double beta decay can be utilized and the operational parameters can be optimized for the energy range of interests. 
\section{0$\mathrm{\nu\beta^+\beta^+}$ sensitivity of a 60 kg LKr detector}
To evaluate the sensitivity to this process and assess the impact of potential backgrounds, a $60\text{ kg}$ detector configuration was simulated using GEANT4 \cite{Agostinelli:2003ed}. The simulated setup consists of a liquid Krypton active volume with a $31.5\text{ cm}$ diameter and a $31.5\text{ cm}$ drift length, enclosed in a $3\text{ cm}$ thick copper vessel. This specific thickness is chosen under the assumption that it sufficiently suppresses the external background, at a level to make it negligible compared to the internal one. The krypton is assumed to have a natural $^{78}\text{Kr}$ abundance of 0.355\%, but with depleted levels of $^{85}\text{Kr}$ (details in appendix \ref{app:Kr85}). Signal events were generated in the active volume by simulating two positrons according to the double beta decay energy spectrum, with the endpoint fixed at the Krypton $Q_\text{Kr}-4m_e$ ($Q_{\text{Kr}} = 2.8463\text{ MeV}$), where $m_e$ is the mass of the electron. The primary background for the $0\nu\beta^+\beta^+$ search arises from high-energy gamma rays interacting via multiple Compton scattering with subsequent bremsstrahlung emission, which might mimic the distinct topology of a double positron decay event. However this gamma must be energetic enough to produce a complex topology in the detector with most probable background events giving rise to pair production events. Thus, only events from $^{208}$Tl and $^{214}$Bi, the most dominant and high energetic gamma background present in the detector material have been simulated. Other lower energy gammas has not been considered since they do not satisfy the requirement of this process. Another source of background is also the 2$\nu$ process which has exactly the same topology of the 0$\nu$ and the events in the tail might enter the region of interest. 

The analysis strategy relies on isolating the main $\beta^+\beta^+$ track and tagging the $511\text{ keV}$ annihilation photons using event reconstruction techniques adapted from Compton imaging \cite{Kim2024}. Specifically, determining the first interaction vertex, scattering angle via a second energy deposition, and energy deposited in the first interaction, under the hypothesis of knowing the interaction point (main $\beta^+\beta^+$ track), is sufficient to tag the production of a 511keV gamma in a given direction. To establish a starting point for this reconstruction (point of the gamma production), each event is required to contain a single track with an energy deposition greater than $550\text{ keV}$ (the main track). This threshold was selected to exceed the photoelectric peak of a $511\text{ keV}$ photon, ensuring an unambiguous origin to evaluate the Compton process and tag the annihilation gamma ray. Consequently, events lacking a track above $550\text{ keV}$, or containing multiple tracks exceeding this threshold, were excluded a priori.

\begin{figure}
    \centering
    \includegraphics[width=.8\linewidth]{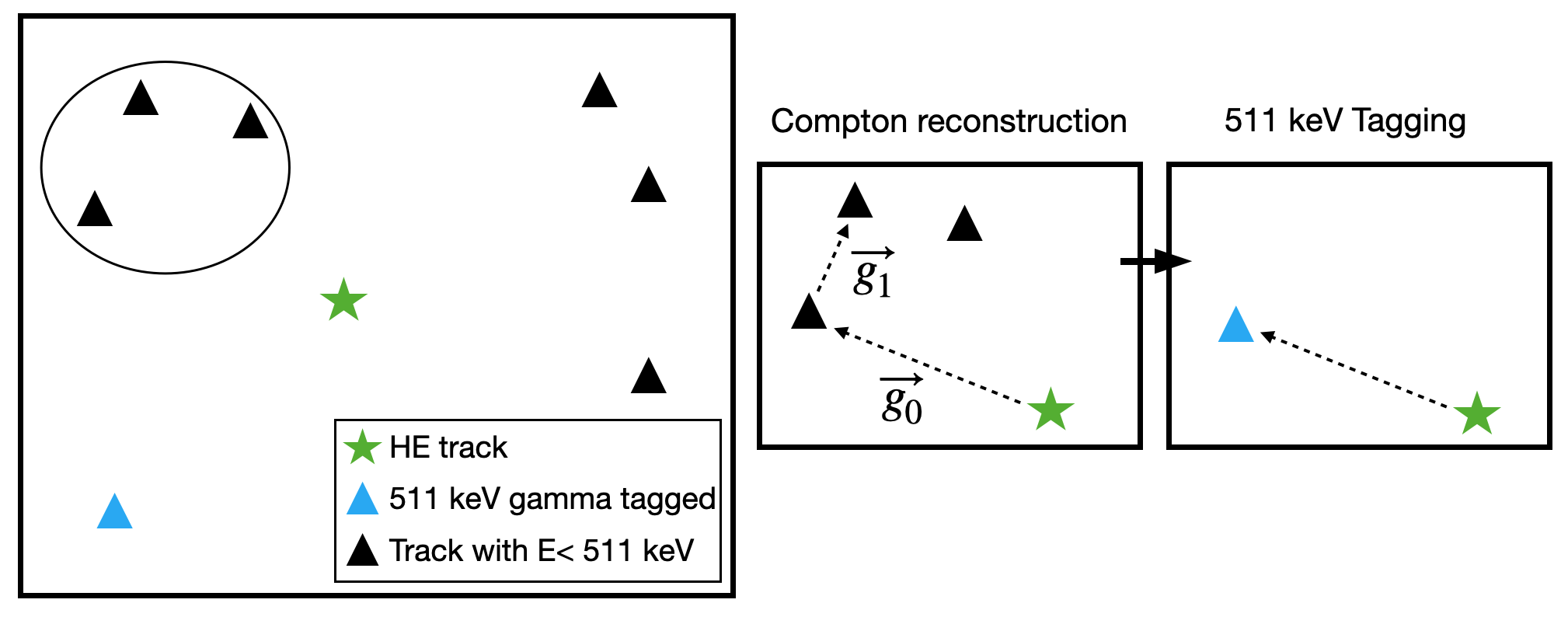}
    \caption{Schematic of the second step in the gamma-tagging sequence for an example of event. Track clusters summing to $511\text{ keV}$ (dark triangles in the circumference) are identified. For each pair of hits, the three-dimensional geometric angle ($\vec{g_0}$  $\vec{g_0}$) is calculated and compared against the expected Compton scattering angle, assuming a $511\text{ keV}$ photon originating from the HE track vertex (green star). Successful candidates are tagged as $511\text{ keV}$ gamma rays at the first interaction point position (cyan triangle).}
    \label{fig:gammareco1}
\end{figure}

\begin{figure}
    \centering
    \includegraphics[width=.8\linewidth]{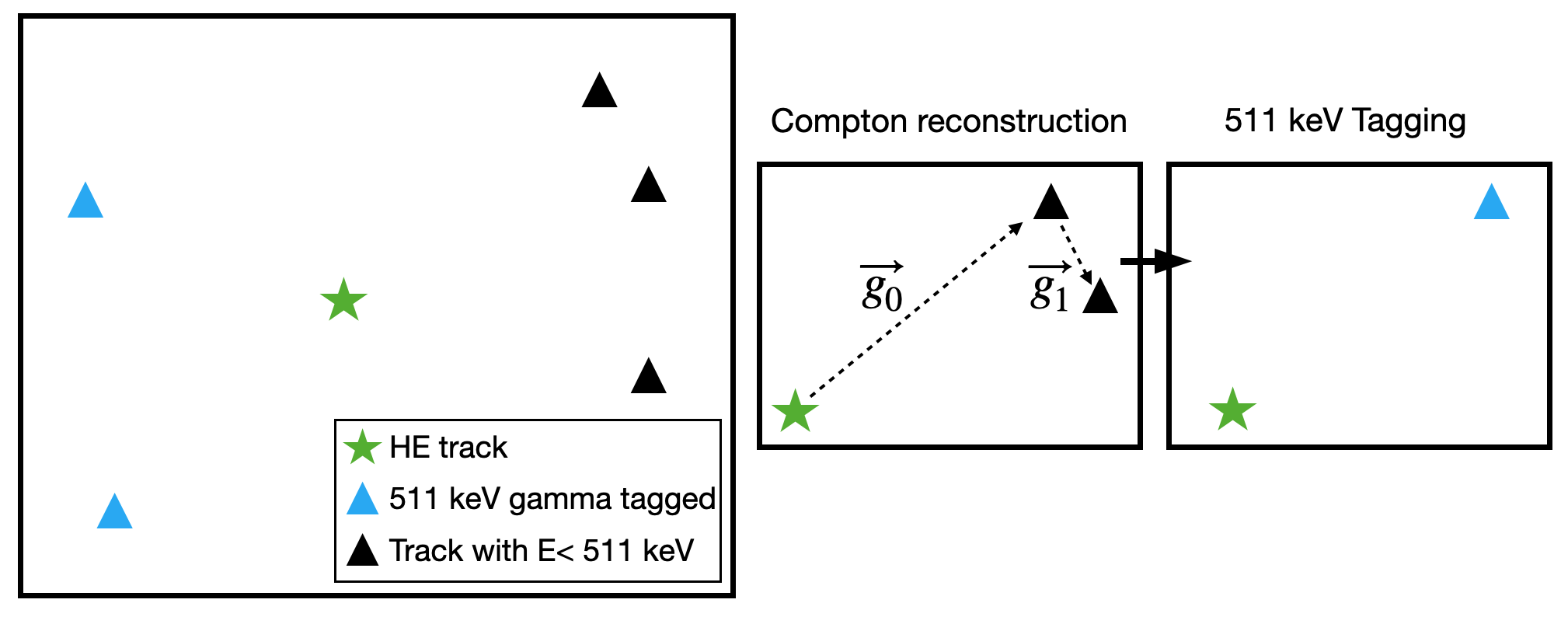}
    \caption{ Schematic of the third step in the gamma-tagging sequence for the same event of Figure \ref{fig:gammareco1}. Each remaining pair of tracks is evaluated for kinematic and geometric consistency with a $511\text{ keV}$ annihilation photon that originates from the HE track vertex and undergoes Compton scattering.}
    \label{fig:gammareco2}
\end{figure}

For each event containing a single selected high-energy (HE) track, the reconstruction sequence consists of three consecutive steps:\\
1) 511 keV tracks are tagged as 511 keV gammas interacting in the track coordinate. \\
2) For each group of tracks with a combined energy deposition equal to $511\text{ keV}$, the hypothesis that they originate from an annihilation photon produced at the high-energy (HE) track vertex is evaluated. For every pair of tracks, the three-dimensional geometric angle $\theta_{geo}$ between them is computed. This value is then compared to the expected Compton scattering angle ($\theta_{\text{ene}}$), which is derived from the deposited energy under the assumption of a $511\text{ keV}$ photon originating from the HE track position.
\begin{equation}
    \theta_{geo} = arccos\left(\frac{\vec{g_0} \cdot \vec{g_1}}{|\vec{g_0}| |\vec{g_1}|}\right) \ \ \ \ \ \ \ \theta_{ene} = arccos\left(1-\frac{m_e}{E_{track}} + \frac{m_e}{511 \mathrm{keV}}\right)
\end{equation}
In these expressions, $\vec{g}_0$ denotes the vector connecting the HE track to the first interaction vertex of the pair, while $\vec{g}_1$ represents the vector connecting the two tracks, oriented from the first to the second. In the kinematic equation, $m_e$ is the electron mass, $E_{\text{track}}$ is the energy deposited at the first interaction point, and the $511\text{ keV}$ term represents the initial energy of the incident gamma ray \cite{Kim2024}.
If the condition $\Delta_\theta = |\theta_{\text{geo}} - \theta_{\text{ene}}| < 0.06$ is satisfied, the $511\text{ keV}$ gamma ray is tagged at the position of the first interaction of the hypothesis tested. In cases where multiple track pairs satisfy this condition, the pair yielding the minimum value of $\Delta_\theta$ is selected. Different values of tolerance have been tested from 0.04 and 0.1 and the variation in the results is not significant. For the calculation, only the track barycenter is used, since it is a parameter significantly more accessible experimentally than the exact track start or end points, which motivates the chosen tolerance threshold. A schematic of this reconstruction process is illustrated in Figure \ref{fig:gammareco1}.\\
3) This procedure is iteratively applied to each remaining pair of unassigned tracks to tag any additional candidate gamma rays, as illustrated in Figure \ref{fig:gammareco2}. Any remaining energy deposits that do not satisfy the tagging criteria are classified as isolated tracks. These tracks represent events where a gamma-ray interacted within the detector volume but failed to deposit its full energy, typically due to the particle escaping the sensitive region before complete absorption.

Subsequently, five distinct event signatures were defined to isolate the signal and discriminate it against the background. To categorize these signatures, spatial collinearity relative to the high-energy track is established: three tracks are considered collinear if they lie along the same linear trajectory. Specifically, a track configuration is defined as collinear with the HE track if the perpendicular distance from the HE track position to the line intersecting the other two tracks is smaller than the length of the HE track's major axis.

Based on these topological criteria, the selected event signatures are defined as follows: \\  \\  

\begin{figure}
    \centering
    \includegraphics[width=1.\linewidth]{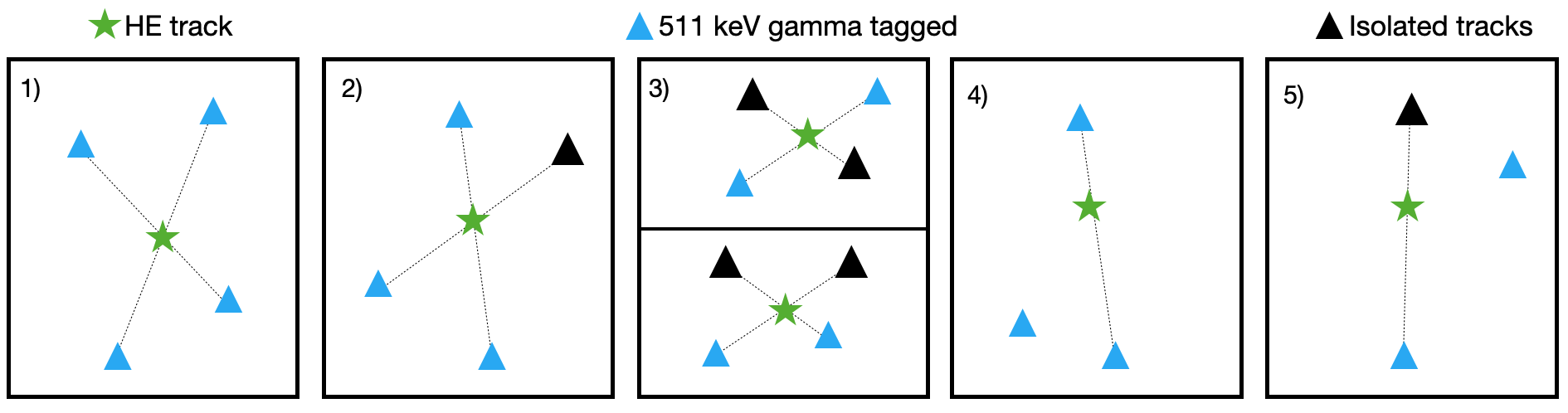}
    \caption{Graphic representation of the 5 signatures used in the analysis. The green stars are the HE track with E$>$550 keV, the cyan triangles are the 511 keV gamma tagged while the gray triangle are the isolated tracks with E$<$511 keV.}
    \label{fig:singatures}
\end{figure}

\begin{itemize}
    \item \textbf{S1} \textbf{($1\,\text{HE} + 4\gamma$):} Consists of one HE track and four tagged $511\,\text{keV}$ photons. The four photons form two distinct pairs, with each pair aligned collinearly with the HE track.
    
    \item \textbf{S2} \textbf{($1\,\text{HE} + 3\gamma + 1\,\text{isolated}$):} Consists of one HE track, three tagged $511\,\text{keV}$ photons, and one isolated track. Two of the photons are collinear with the HE track, while the third photon is collinear with the isolated track and the HE track.
    
    \item \textbf{S3} \textbf{($1\,\text{HE} + 2\gamma + 2\,\text{isolated}$):} Consists of one HE track, two tagged $511\,\text{keV}$ photons, and two isolated tracks. The four non-HE objects form two distinct pairs, with both pairs aligned collinearly with the HE track.
    
    \item \textbf{S4} \textbf{($1\,\text{HE} + 3\gamma$):} Consists of one HE track and three tagged $511\,\text{keV}$ photons. Two of the photons are collinear with the HE track, while the third is unaligned.
    
    \item \textbf{S5} \textbf{($1\,\text{HE} + 2\gamma + 1\,\text{isolated}$):} Consists of one HE track, two tagged $511\,\text{keV}$ photons, and one isolated track. The two photons are not collinear with each other, but one photon is collinear with the isolated track.
\end{itemize}
A graphic representation of these 5 signature used in the analysis is shown in Figure \ref{fig:singatures}. 

A Monte Carlo simulation in GEANT4 of $100,000$ signal events, $2.0 \times 10^{7}$ $^{208}\text{Tl}$ decays, and $4.0 \times 10^{7}$ $^{214}\text{Bi}$ decays. The detector setup has a cylindrical liquid chamber ($31.5\text{ cm}$ diameter, $31.5\text{ cm}$ drift length) surrounded by a $3\text{ cm}$ thick copper shield. Based on the copper radioactivity values from \cite{2026}, the total radioactivity for the whole vessel is $0.4 \times 10^{-3}\text{ Bq}$ for $^{208}\text{Tl}$ and $2.5 \times 10^{-3}\text{ Bq}$ for $^{214}\text{Bi}$. These simulated events consists of an effective simulated time of $1520\text{ years}$ for $^{208}\text{Tl}$ and $478\text{ years}$ for $^{214}\text{Bi}$, respectively.

For each signature, the selection efficiency and the expected number of background events per 10 years of exposure are reported in Table \ref{tab:eff_tab}.
\begin{table}[]
\centering
\begin{tabular}{|c|c|c|c|c|}\hline
Signature & $0\nu$ S.E. (\%) & $2\nu$ S.E. (\%) & $^{208}$Tl Bkg rate (per 10 y) & $^{214}$Bi Bkg rate (per 10 y) \\ \hline
1&6.87    &0.26      &$<$0.0147       &$<$0.042          \\
2&4.54     &0.209          &0.0861      &$<$0.042        \\
3&1.00    &0.07       &0.651      &0.126      \\
4&15.07  &2.51             &4.956      &27.12        \\
5&6.97   &0.96           &13.608      &41.60     \\ \hline 
\end{tabular}
\caption{Selection efficiencies and expected background rates for the five event signatures defined in this work. Upper limits correspond to a $90\%$ confidence level, determined from the zero observed events in the simulation.}
\label{tab:eff_tab}
\end{table}

\begin{figure}
    \centering
    \includegraphics[width=1.05\linewidth]{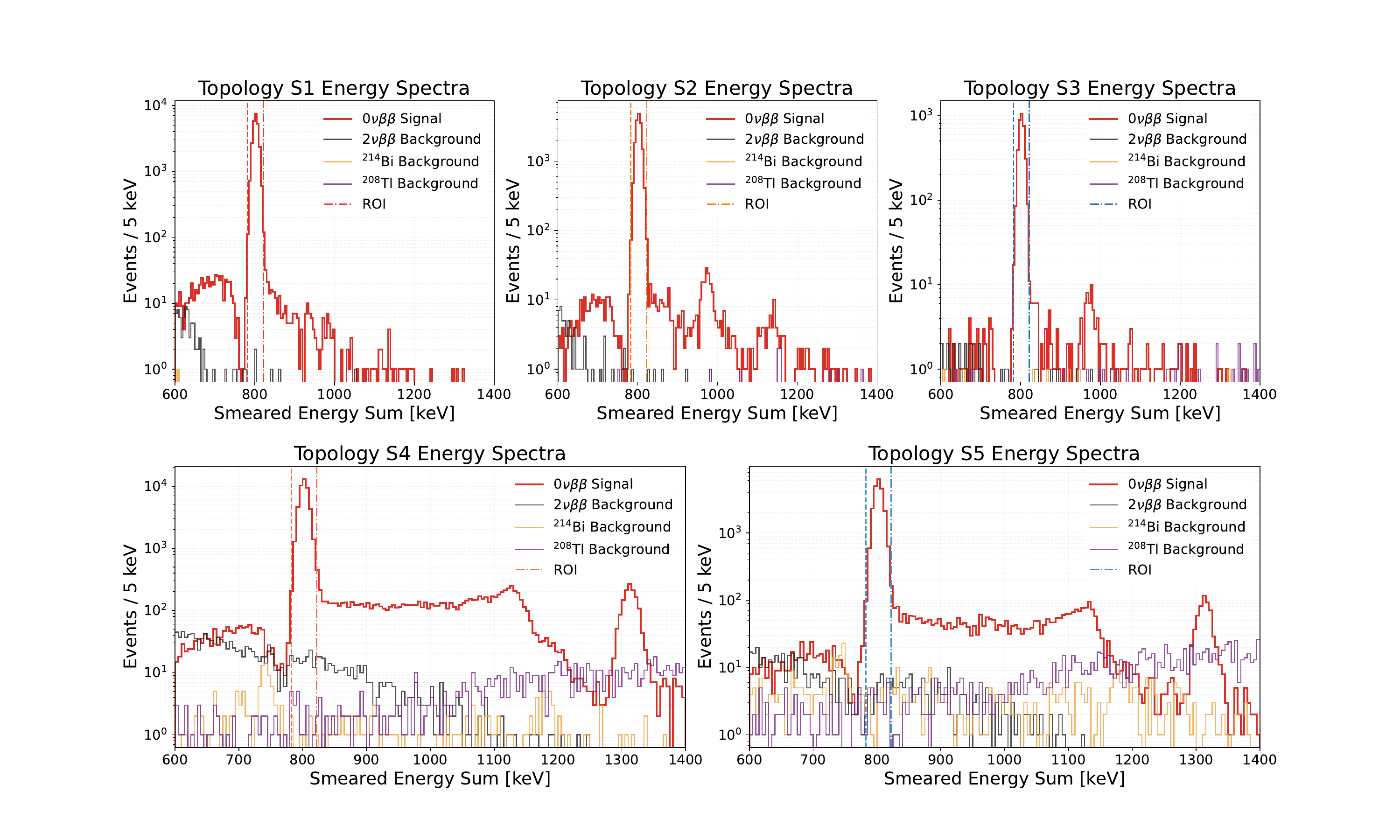}
    \caption{Spectra of the HE track for the different topological selection and the cut in the region of interest (ROI).}
    \label{fig:onuspectra}
\end{figure}

A further selection cut is applied to events falling within the region of interest (ROI). Assuming a $1\%$ energy resolution ($\sigma$) at $802\text{ keV}$ \cite{2023JInst..18C4007P}, the expected energy for the double positron track, the energy distribution of the HE track was computed, and a $\pm3\sigma$ cut was applied around the peak value. The resulting energy distributions for the different components and for the different signatures are shown in Figure \ref{fig:onuspectra} together with the ROI selection.

The $511\text{ keV}$ gamma interaction spectrum observed to the right of the main peak in topologies 4 and 5 is due to the spatial merging of the gamma ray's Compton or photoelectric track with the primary HE track.
This selection in energy applies a further acceptance efficiency in signal and background shown in Table\ref{tab:selection_efficiencies_full}.

\begin{table}[]
\centering
\begin{tabular}{|c|c|c|c|c|}
\hline
Signature & Eff. $0\nu$ (\%) & Eff. $2\nu$ (\%) & $^{208}$Tl Bkg rate (per 10 y) & $^{214}$Bi Bkg rate (per 10 y) \\ \hline
1 & $6.38$ & $0.0026$ & $<0.01$ & $<0.04$ \\
2 & $4.74$ & $0.0013$ & $<0.01$ & $<0.04$ \\
3 & $1.24$ & $0.0001$ & $0.06$ & $<0.04$ \\
4 & $11.5$ & $0.16$ & $0.19$ & $0.14$ \\
5 & $6.26$ & $0.048$ & $0.42$ & $0.18$ \\ \hline
\end{tabular}
\caption{Total selection efficiencies ($\%$) for the signal ($0\nu\beta\beta$) and major background components ($2\nu\beta\beta$, $^{214}\mathrm{Bi}$, and $^{208}\mathrm{Tl}$) across the evaluated topology signatures within the defined ROI. Also here upper limits correspond to a 90\% confidence level, determined from the zero observed events in the simulation.}
\label{tab:selection_efficiencies_full}
\end{table}

\begin{figure}
    \centering
    \includegraphics[width=0.7\linewidth]{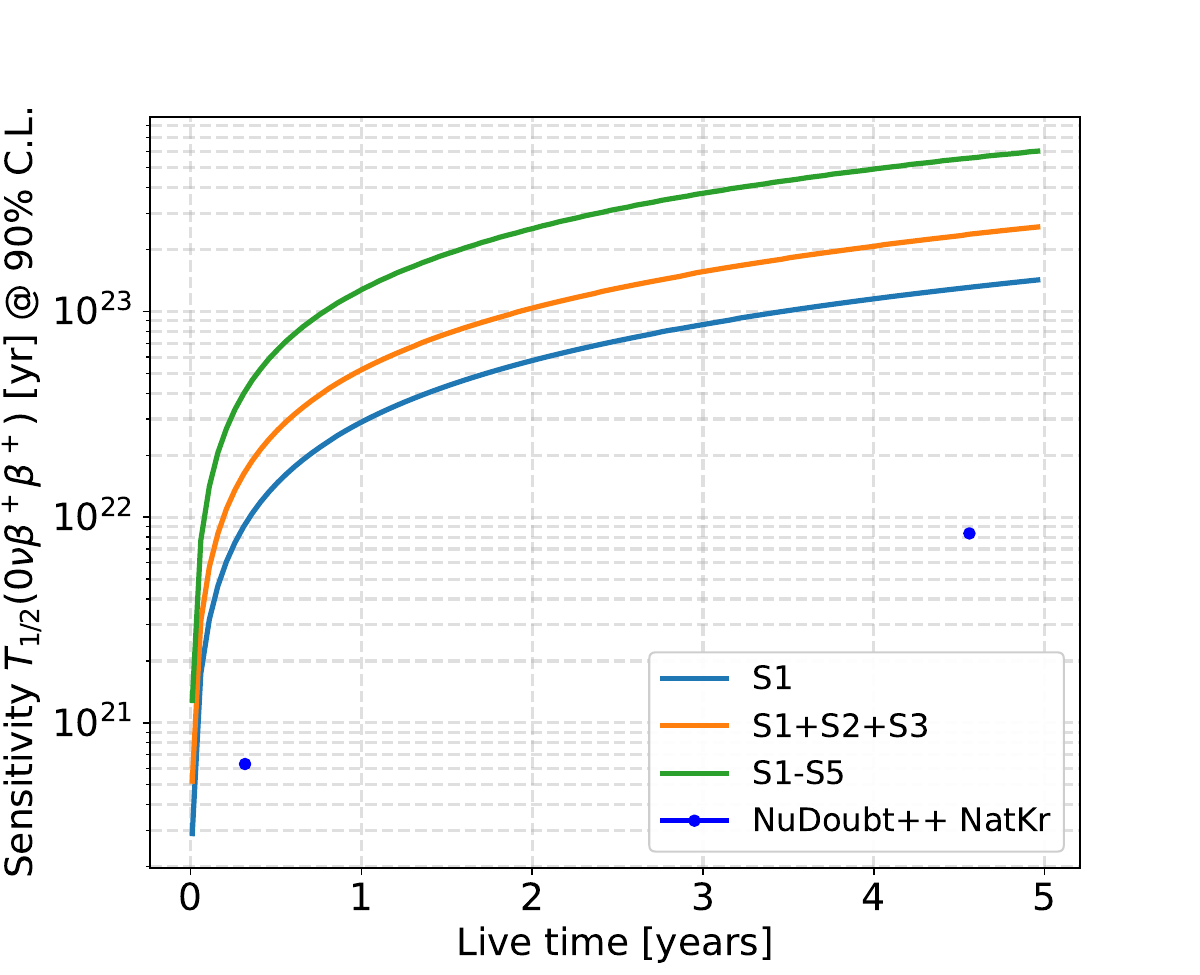}
    \caption{Projected sensitivity curves for the $0\nu\beta^+\beta^+$ decay process using a $60\text{ kg}$ natural Krypton target. Each curve displays the total sensitivity achieved by combining the different signature. The different signature used are reported in the legend, together with the sensitivity projection form \cite{B_hles_2025}.}
    \label{fig:sens_curve}
\end{figure}

Assuming a lifetime of $10^{24}$ an activity for the 2$\nu$ process of 0.28 ev/y can be assumed in a 60 kg detector. Multiplying it by the selection efficiency values in Table\ref{tab:selection_efficiencies_full} also this background will be at the level of $<<$1 ev/ (10 y) leading to a virtually background free search for the 0$\nu$ process. 
For the sensitivity calculation the different signatures have been statistically
treated as independent experiments, each following an independent Poisson
distribution for its expected signal and background,
\begin{equation}
    P(n_i \mid \gamma) = \frac{\lambda_i(\gamma)^{n_i}\, e^{-\lambda_i(\gamma)}}{n_i!},
    \qquad
    \lambda_i(\gamma) = K\,\varepsilon_i\,\gamma + b_i ,
\end{equation}
where $\gamma \equiv 1/T_{1/2}$ is the decay rate, $\varepsilon_i$ and $b_i$ are the
efficiency and expected background rate of signature $i$, and
$K = \ln2 \cdot N_{target} \cdot t$. The combined sensitivity was obtained via a
multichannel generalization of the Feldman-Cousins unified approach, with the
joint likelihood built as the product of the per-signature Poisson
probabilities,
\begin{equation}
    P(\vec n \mid \gamma) = \prod_{i=1}^{C} P(n_i \mid \gamma),
\end{equation}
and acceptance regions constructed through the likelihood-ratio ordering
principle applied to the vector of channel counts $\vec n = (n_1,\dots,n_C)$,
ranking outcomes by
\begin{equation}
    R(\vec n,\gamma) = \frac{P(\vec n \mid \gamma)}{P(\vec n \mid \hat\gamma(\vec n))},
\end{equation}
with $\hat\gamma(\vec n) \geq 0$ the physically-allowed best-fit rate for that
outcome. The sensitivity is defined as the average 90\% C.L. upper limit on the decay rate over the ensemble of background-only outcomes,
\begin{equation}
    \langle \Gamma_{\rm up} \rangle = \sum_{\vec n} P(\vec n \mid \gamma{=}0)\, U(\vec n),
    \qquad
    T_{1/2}^{\rm sens} = \frac{1}{\langle \Gamma_{\rm up} \rangle},
\end{equation}
where $U(\vec n)$ is the Feldman-Cousins upper limit on $\gamma$ for the count vector $\vec n$, following \cite{GomezCadenas2011}. The same procedure is applied to individual signatures and their combinations, and repeated over a range of live times $t$ to project the resulting sensitivity as a function of exposure for each configuration.
Under this construction the sensitivity curves are shown in Figure \ref{fig:sens_curve} for the single S1 signature (the cleanest), the signatures S1, S2 and S3 combined (the ones with extremely low background) and all 5 signatures combined, and compared with current theoretical prediction \cite{B_hles_2025} under the same exposure conditions.

A further analysis utilizing the Monte Carlo truth information on the gammas was performed to assess the gamma tagging efficiency. In this evaluation, a gamma ray is considered taggable if it undergoes a photoelectric interaction after, interacts at least twice within the TPC via Compton scattering, or interacts and subsequently escapes to form a cluster (considered as isolated track in this case). This truth-level study shows that an efficiency of $17\%$ can be reached for the first three signatures, with a total efficiency of $40\%$ for the 5 signature for this detector geometry. This indicates that there is still space for algorithm improvement to enhance the sensitivity.

\section{2$\mathrm{\nu\beta^+\beta^+}$ sensitivity of a 60 kg LKr detector}

\begin{figure}
    \centering
    \includegraphics[width=0.7\linewidth]{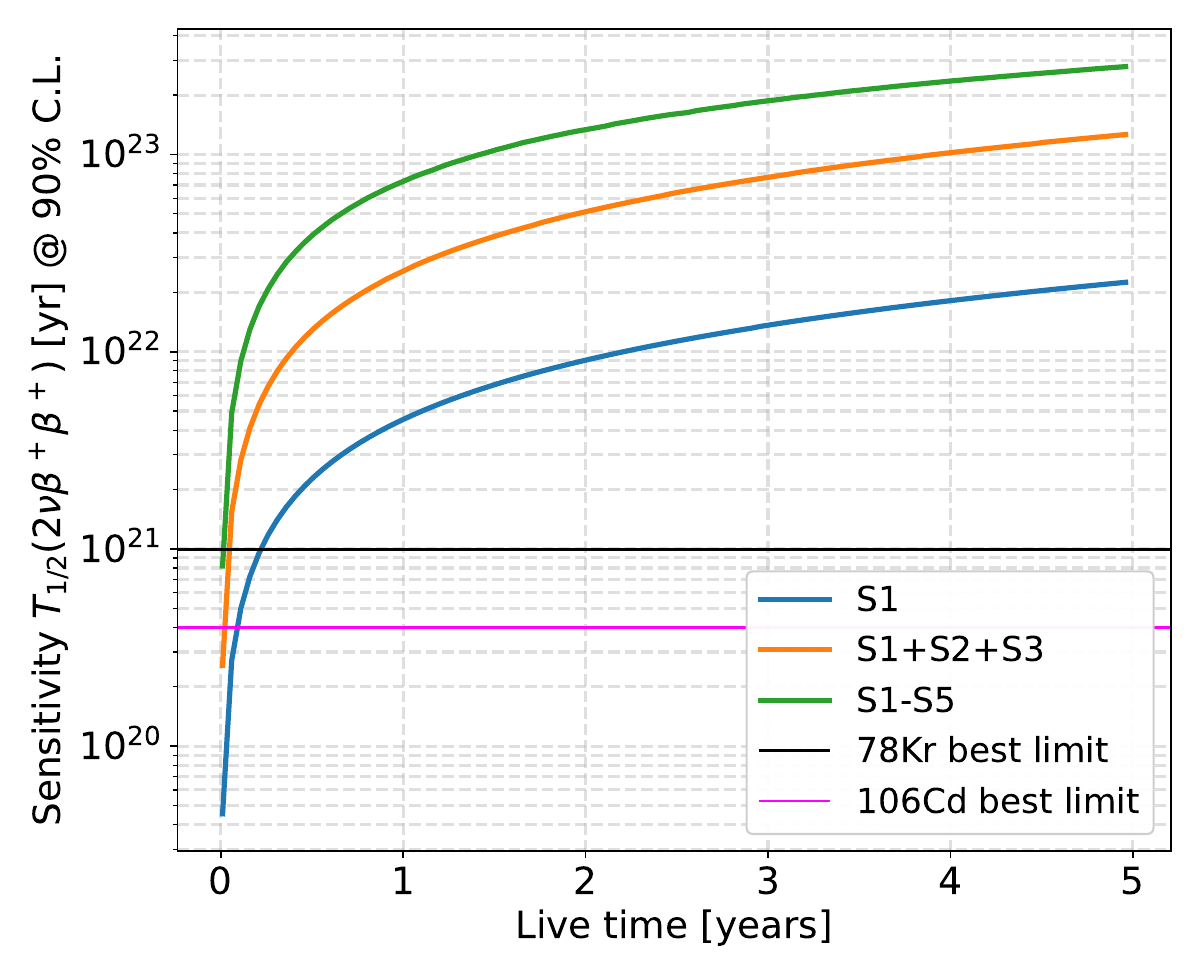}
    \caption{Projected sensitivity curves for the $2\nu\beta^+\beta^+$ decay process using a $60\text{ kg}$ natural Krypton target. Each curve displays the cumulative sensitivity achieved by combining successive signature efficiencies. The different signature used are reported in the legend. The two horizontal limits correspond to the current world-best limits established for $^{106}\text{Cd}$ \cite{Belli_1999} and $^{78}\text{Kr}$ \cite{SAENZ1994363}.}
    \label{fig:sens_2nu}
\end{figure}

The sensitivity was also investigated for the corresponding $2\nu\beta^+\beta^+$ process. For this purpose, the only modification to the algorithm was made in the identification of the main track (HE track) of the event. Because the $2\nu$ energy spectrum is a continuum with an endpoint at $802\text{ keV}$, requiring the HE track to have an energy $E > 550\text{ keV}$ would strongly reduce the selection efficiency. On the other side, the energy can no longer be used to isolate the double positron track, necessitating a dedicated track-identification algorithm. Even if the $511\text{ keV}$ gamma rays interact entirely via Compton scattering, the characteristic collinear behavior between pairs of tracks must still be satisfied. Therefore, if no track with $E > 550\text{ keV}$ is present in the event, the collinearity is evaluated for every track pair. If a given track lies within $1\text{ mm}$ of the line connecting two other tracks, it receives a count. The track with the highest cumulative count is then selected as the primary track. Once the main track is identified, the rest of the analysis proceeds identically to the $0\nu$ study. Following this selection, an energy window of $100\text{ keV} < E < 850\text{ keV}$ is applied to the main track to further suppress background.
After the selection the efficiency on the resulting signal efficiency and the background rate is reported in Table \ref{tab:eff2nu}.
\begin{table}[]
\centering
\begin{tabular}{|c|c|c|c|}
\hline
Signature & Eff. 2$\nu$ (\%) & $^{208}$Tl Bkg rate (per 10 y) & $^{214}$Bi Bkg rate (per 10 y) \\ \hline
1         & 1.04 & $<$0.01  & $<$0.004                       \\
2         & 2.46 & 0.006  & 0.041                          \\
3         & 2.39 & 0.019  & 0.083                          \\
4         & 5.27 & 0.25  & 1.20                           \\
5         & 7.61 & 1.06  & 5.20       \\     \hline              
\end{tabular}
\caption{Selection efficiencies and expected background rates for the five event signatures defined in this work. Upper limits correspond to a $90\%$ confidence level, determined from the zero observed events in the simulation.}
\label{tab:eff2nu}
\end{table}
The sensitivity for the $2\nu$ case is calculated as the one for the $0\nu$ and is show in Figure \ref{fig:sens_2nu} for the same signature combinations. The current two best limits for $^{106}\text{Cd}$ \cite{Belli_1999} and $^{78}\text{Kr}$ \cite{SAENZ1994363} are also shown in the legend. 
\section{Conclusions}
While main detectors searching for double beta decay with extremely low background and exceptional energy resolution suffer for scalability due to complexity and high cost, and easier to scale experiment like large TPC or crystals can scale dominated by background on larger scale version \cite{GomezCadenas2011}, such a detector would feature the easy scalability typical of the large TPC while still maintaining no background for the extremely clear topology of the decay. 

The search for double positron decay offers a clear, unique way to investigate whether neutrinos are Majorana particles. This is due to the very distinct signature of the decay, which produces four $511\text{ keV}$ gamma rays. Because of this multi-particle signature, a Time Projection Chamber is ideal for this research since it naturally provides three-dimensional imaging. Among the isotopes that undergo $\beta^+\beta^+$ decay and can serve as a detector gas, Krypton stands out for its low cost and high performance.

A Krypton TPC operates at a density that is low enough to separate the gamma rays from the initial positron annihilation point, but high enough to stop and contain the photons within the active volume. Additionally, electron diffusion is limited to just $1\text{ mm}$ over a $50\text{ cm}$ drift distance, which allows for excellent track separation when using a readout grid with $\sim$ $3\text{ mm}$ pitch.

A first simulation with natural Kr shows that a detector with a 31.5 cm diameter and 31.5 cm drift length containing only 60 kg of natural Kr can reach a sensitivity of $> 10^{23}$ y in 1.5 years for both the $0\nu$ and $2\nu$ processes. Furthermore, the simulation demonstrates the exceptionally high rejection power of the topological selection analysis, which reduces background due to $^{208}$Tl and $^{214}$Bi to below 1 count per 10 years for most signatures, and maintains them at fewer than 1 event per year even in the worst-case scenarios. $^{40}$K has been additionally analyzed but not accounted for since after the topological selection in the whole energy range only 20 events out of $4\cdot 10^7$ passed the topological selection with a negligible impact on the analysis.  
At the moment, simple two-point Compton algorithms have been used to tag the 511 keV gammas, and a more refined version that accounts for multiple interactions or machine learning techniques might improve the gamma-tagging efficiency \cite{Kim2024} both increasing the signal efficiency, and the background rejection in improving the rejection of false 511 keV gamma tagging.

A larger version of the detector on the ton scale might, in addition to exposing more mass, increase the gamma detection efficiency due to better containment leading to a full reconstruction in the cases in which one gamma is lost because it escapes the detector sensitive volume. Signatures S1, S2, and S3, which require four tracks in addition to the primary track, provide significantly better background rejection than the three-track signatures, S4 and S5. Consequently, these first three signatures will remain robust even in larger-scale detectors where higher background rates are expected. In a larger detector, the efficiency of signatures S2–S5 is expected to shift in favor of signature S1, due to improved gamma containment, ensuring the scalability of this technique to larger scale detectors. In this regime, the primary limiting factor becomes gamma rays interacting in close proximity to their production vertex.
Thus, a ton-scale detector, benefiting from an increased detection efficiency, could effectively reach sensitivities of the order of $10^{27}$ y in 3 years even with natural Kr, which has an isotopic abundance of 0.35\%. Furthermore, a ton-scale detector enriched with $^{78}\text{Kr}$ could effectively reach sensitivities on the level of $10^{29}-10^{30}$ y over time.  


\section*{Acknowledgments}
The authors want to thank the European Research Council (ERC) under Grant Agreement No. 951281-BOLD, 101039048-GanESS, and 101165559-COLINA, the Spanish State Research Agency through the Plan Nacional (PID/Proyectos de Generación de Conocimiento) PID2021-125475NB, and the Ramón y Cajal program through Grant RYC2023-045436-I, funded by MICIU/AEI/10.13039/501100011033 and FSE+. Many thanks also to Roberto Soleti for his valuable suggestions.
\newpage
\appendix
\section{The $^{85}$Kr background in a time projection chamber}
\label{app:Kr85}
The primary limiting factor of this technology is the presence of the radioactive isotope $^{85}\text{Kr}$ in natural krypton extracted from the atmosphere, which currently remains the primary commercial extraction method. $^{85}\text{Kr}$ is a radioactive isotope of the noble gas krypton that currently exists in the Earth’s atmosphere with a concentration of about \(0.087 \text{ ppm}\). It possesses a half-life of $10.74\text{ years}$ and decays into stable rubidium-85 ($^{85}\text{Rb}$) via two beta-decay branches. The dominant transition exhibits a maximum beta energy of $0.678\text{ MeV}$ with a $99.56\%$ branching ratio, while the secondary channel has a maximum energy of $0.173\text{ MeV}$ and a branching ratio of $0.44\%$ \cite{BOLLHOFER20197}. The presence of atmospheric $^{85}\text{Kr}$ is largely anthropogenic in origin. While a negligible fraction is produced naturally via cosmic-ray neutron activation of stable atmospheric krypton, the dominant sources are nuclear fission and the reprocessing of spent nuclear fuel. Since the mid-20th century up to the end of 2009 5500 PBq of $^{85}\text{Kr}$ has been released in the atmosphere \cite{AHLSWEDE201334}. These activities have elevated the global atmospheric concentration of $^{85}\text{Kr}$ to a stable abundance of $10^{-11}$ nowadays \cite{Yangetal2013}.

The current $^{85}\text{Kr}$ abundance corresponds to a specific activity of approximately 147 kBq/kg in natural krypton. For the TPC design considered in this work, which features a 60 kg active mass and a 31.5 cm drift length, assuming a drift velocity of $2 \times 10^5\text{ cm/s}$ \cite{PhysRevA.14.438} yields a maximum drift time of $157.5\ \mu\text{s}$. Consequently, a single data acquisition window would contain an average of 1,389 overlapping background events. Such an extremely high pile-up rate would make operating the TPC practically unfeasible under standard conditions. Nevertheless, several physical factors saves the viability of a krypton-based TPC. Most notably, the $^{85}\text{Kr}$ contaminant is separated by seven atomic mass units from the target isotope, $^{78}\text{Kr}$. Positioned between them is a large mass bulk of stable isotopes, $^{82}\text{Kr}$, $^{83}\text{Kr}$, and $^{84}\text{Kr}$, which collectively constitute approximately 80\% of natural krypton's abundance. 

This mass difference can be easily exploited for separation with the use of centrifuges. Inside a high-speed rotating centrifuge, the centrifugal force drives the heavier isotopes toward the outer wall, while the lighter isotopes remain closer to the central axis. By making a mass cut within the large bulk of stable isotopes, the heavy radioactive $^{85}\text{Kr}$ can be cleanly isolated and discarded along with a fraction of the gas. This process leaves behind a low-background krypton depleted in $^{85}\text{Kr}$ which will also be as a consequence enriched in $^{78}\text{Kr}$ from the separation. While a precise purification factor has not yet been experimentally demonstrated, the expected isotopic purity should be exceptionally high cutting in the bulk. The feasibility of the cut in the bulk instead of the precise separation of the $^{85}\text{Kr}$ isotope from the $^{84}\text{Kr}$ drastically cut costs and technical complexity of the process. Anyway the so called "light cut" of the natural Kr centrifugate is commercially available.

Other methods, including cryogenic distillation, thermal diffusion, and underground sourcing, after an investigation has been evaluated as impractical due to small krypton isotopes volatility difference, extremely long processing times, and low atmospheric abundance. However, for the feasibility point of view in \cite{kuzminov1992radioactive} (1992) it is demonstrated that a large quantity of 98\% enriched $^{78}$Kr with a content of $^{85}\text{Kr}$ 4$\cdot$10$^3$ times less than natural Kr has been produced. This demonstrates that the enrichment in $^{78}$Kr and the depletion in $^{85}\text{Kr}$ by orders of magnitude is industrially feasible.

From the point of view of the remaining isotope, several strategies can be employed to suppress this background. A pileup event of a beta decay of $^{85}\text{Kr}$ in a $\beta^+\beta^+$ event would produce a distinct second primary scintillation signal that should be clearly visible in the waveform. Furthermore, although not yet modeled in this work, such a pileup event is highly unlikely to mimic the kinematics characterized by a high-energy track and two pairs of collinear tracks and to participating in the mimicking the kinematic of a 511 gamma emitted from the main track. In larger detectors, spatial containment can also be exploited: given an interaction length of $\sim$5 cm in liquid krypton (LKr), 95\% of the 511 keV gammas will interact within 15 cm. Consequently, the signal event is typically contained within a 30 cm diameter sphere, allowing any coincident pileup interactions occurring outside this volume to be identified and discarded. Finally, from a trigger perspective, setting the S1 scintillation threshold above 1 MeV (with 100\% signal efficiency) would entirely prevent triggering on these lower-energy $^{85}\text{Kr}$ beta decays.

\newpage

\printbibliography 

\end{document}